\documentclass[sigconf]{acmart}

\usepackage{tikz}
\usepackage[english]{babel}
\usepackage[utf8]{inputenc}
\usepackage{pifont}
\usepackage{caption}
\usepackage{subcaption}
\usepackage{rotating}
\usepackage{multirow}
\usepackage{mdframed}
\usepackage[most]{tcolorbox}
\newcommand\RotText[1]{\rotatebox{90}{\parbox{1.9 cm}{\raggedright#1}}}

\newcommand*\OK{\ding{51}}
\newcommand{\xmark}{\ding{55}}%

\mdfdefinestyle{MyFrame}{%
    linecolor=black,
    outerlinewidth=2pt,
    innertopmargin=4pt,
    innerbottommargin=4pt,
    innerrightmargin=4pt,
    innerleftmargin=4pt,
        leftmargin = 4pt,
        rightmargin = 4pt
        }

\usetikzlibrary{arrows,decorations.pathmorphing,backgrounds,fit,positioning,shapes.symbols,chains}
\newcommand\tikzmark[2]{%
\tikz[remember picture,overlay] 
\node[inner sep=1pt,outer sep=2pt] (#1){#2};%
}
\newcommand\link[2]{%
\begin{tikzpicture}[remember picture, overlay, >=stealth, shorten >= 2pt]
  \draw[->] (#1.east) to  (#2.west);
\end{tikzpicture}%
}

\begin{document}
\fancyhead{}

\title{A Taxonomy for Contrasting Industrial Control Systems Asset Discovery Tools}

\author{Emmanouil Samanis}
\affiliation{%
  \institution{Bristol Cyber Security Group, University of Bristol}
  \city{Bristol}
  \country{UK}}
\email{manolis.samanis@bristol.ac.uk}

\author{Joseph Gardiner}
\affiliation{%
  \institution{Bristol Cyber Security Group, University of Bristol}
  \city{Bristol}
  \country{UK}}
\email{joe.gardiner@bristol.ac.uk}

\author{Awais Rashid}
\affiliation{%
  \institution{Bristol Cyber Security Group, University of Bristol}
  \city{Bristol}
  \country{UK}}
\email{awais.rashid@bristol.ac.uk}

\begin{abstract}
Asset scanning and discovery is the first and foremost step for organizations to understand what assets they have and what to protect. There is currently a plethora of free and commercial asset scanning tools specializing in identifying assets in industrial control systems (ICS). However, there is little information available on their comparative capabilities and how their respective features contrast. Nor is it clear to what depth of scanning these tools can reach and whether they are fit-for-purpose in a scaled industrial network architecture. We provide the first systematic feature comparison of free-to-use asset scanning tools on the basis of an ICS scanning taxonomy that we propose. Based on the taxonomy, we investigate scanning depths reached by the tools' features and validate our investigation through experimentation on Siemens, Schneider Electric, and Allen Bradley devices in a testbed environment.
\end{abstract}

\maketitle

\keywords{Industrial Control Systems; ICS Security; Asset scanning; Critical infrastructure; SCADA; Operational Technology; Cyber-physical security; Port scanning; Programmable logic controllers}

\section{Introduction}\label{sec:I.intro}
Asset scanning is the process of discovering and collecting information about all physical and logical assets connected to a network, often implemented with the use of scanning tools. Recommended as a practice in the \textit{identify} stage of the NIST Framework~\cite{DBLP:journals/Barrett20}, it is a key element of risk assessments -- discover which assets are connected to the organization's network in order to identify known vulnerabilities and put in place mitigating actions. It is also the stepping stone for a defense strategy. Security teams require effective asset scanning tools to understand the potential attack surface as new devices connect to the network or, existing ones are updated or modified in response to particular business needs. It is a well-established practice in IT networks with many commercial and free/open-source tools available for the purpose. In recent years, as industrial control systems (ICS) responsible for running critical national infrastructures, such as water treatment, power systems, manufacturing and other industrial environments have expanded in use, specialist asset scanning tools have emerged for this new setting. This is particularly driven by increasing network connectivity of such systems, including to the wider internet (traditionally, such systems were air-gapped for security). Some of these are bespoke solutions for Operational Technology (OT) -- a term often used to describe the controllers, sensors and actuators deployed in industrial settings. 

Whilst a traditional IT system is largely responsible for handling the flow of information, an OT system is responsible for controlling and monitoring a physical, real-world process, with potentially catastrophic consequences if something goes wrong. OT systems can be vast and complex, consisting of thousands of specialized devices and numerous pieces of software, used for monitoring and interacting with such devices that grow over the years as new equipment is added. For these complex critical infrastructure architectures, asset scanning tools not only help with the auditing process for the industrial devices and identify where security improvements are needed (e.g., firmware updates) but also assist in the commissioning process of new equipment \cite{DBLP:conf/acsc/RodofileRF16}. 

Reconnaissance is an important step for an attacker as described in ATT\&CK matrix for enterprise which includes a knowledge base of adversary tactics and techniques~\cite{DBLP:enterprise/mitre}.This step is the first in the planning phase of the ICS cyber kill chain, where adversaries actively or passively gather information to identify potential targets or exfiltrate abundant information (device properties or vulnerabilities)~\cite{DBLP:books/sp/19/GroobyDD19a}. Asset scanning tactics and techniques are included in the discovery stage of the MITRE ATT\&CK matrix for ICS, which shows, at various stages, the potential actions of an attacker's intrusion into an ICS network \cite{DBLP:collaborate/mitre}.

Furthermore, an ICS environment will often contain legacy devices (ICS equipment is designed to operate for decades with minimal interruption), which can experience issues when exposed to scanning activities, in particular, the packet-heavy approaches that are used in IT environments. For example, an older programmable logic controller (PLC) with a low powered CPU or poorly implemented network stack could be overloaded by the high rate port and service scanning behavior from a common tool such as Nmap, if used without due regard to such considerations~\cite{DBLP:journals/Greenbone18}. Further, the use of real-time industrial communication protocols that expect a steady stream of data from ICS devices can be interrupted by the heavy network traffic communications caused by active scanning of such devices. Due to these issues, as well as the use of specialized communication protocols used in ICS settings, eg., S7comm, DNP3, Profinet, etc., asset scanning for ICS cannot simply be achieved by transposing IT asset scanning tools to OT environments~\cite{Stouffer2011}. This is particularly critical because the requirements in ICS are different than IT systems and focus on safety, reliability, robustness, and maintainability \cite{DBLP:journals/compsec/CherdantsevaBBE16}. A violation of these attributes could result in human causalities, physical damage to the industrial process or large scale societal disruption of critical services. Therefore, asset scanning tools must not only support the specialized equipment and protocols but also account for critical properties such as safety, reliability and real-time requirements.

Unlike IT environments where comparative analyses have been conducted \cite{DBLP:journals/comsur/Bou-HarbDA14}, and despite the growing number of ICS asset scanning tools, there lacks an analytic framework to help understand and contrast the full spectrum of asset scanning techniques and methods for OT environments. The lack of such a framework -- and systematic investigations based on such a framework -- makes it difficult for asset owners to decide which tools may be suitable to their particular ICS environment, whether the suitable tools afford the required scanning depths and other features such as active or passive scanning (with the former having potential for disruption to legacy environments). This paper addresses this gap by:
\begin{itemize}
\item Introducing a taxonomy for ICS asset scanning tools, their various features and scanning depth levels. To our knowledge, ours is the first taxonomy proposed to date that enables systematic mapping, characterization and classification of the features offered by ICS asset scanning tools. The taxonomy -- and the capacity to contrast tools afforded by it -- will enable users to garner a more objective understanding of the potential applicability of tools within their infrastructures and suitability to requirements and safety considerations (e.g., potential disruption of industrial processes due to active scanning);

\item Contrasting the features and functionality -- as depicted in their documentation or implementation -- of twenty eight free-to-use and shareware tools on the basis of the taxonomy.  We choose free-to-use tools mainly because they are a better fit for small infrastructure operators who have limited resources. As investment in expensive commercial tools may not be an option for such resource stretched smaller companies~\cite{DBLP:power/suppliers};

\item Evaluating these features in a realistic ICS testbed (See Section ~\ref{sec:lab}) to both establish the effectiveness of the tools' features and investigate their safe operation within real industrial networks. The experimental evaluation does not only provide an insight into the scanning depth and quality of each tool but also the risk it may pose due to active scanning and the effects of such scans on the production.
\end{itemize}
To our knowledge, we are the first to propose a taxonomy to compare and contrast ICS asset scanning tools and use it as a basis to compare (both analytically and experimentally) twenty-eight tools. Our analysis also provides a first baseline comparison of the features of the 28 tools we studied, enabling future analysis with regard to the baseline as these tools evolve or new tools come on the scene (as well as comparisons with commercial offerings in future studies).

\section{ICS system architecture}\label{sec:II.architecture}

In Figure~\ref{fig:ICS/SCADA Architecture} we provide an example of a “typical” ICS deployment architecture. Often, the Purdue model is used to hierarchically categorize the architecture of an ICS~\cite{DBLP:books/sp/Flaus19}, and does indeed map to our architecture. However, we provide a much more detailed and realistic representation of the specific devices and systems within a typical ICS deployment compared to the high-level view of the Purdue model. Our example architecture highlights the challenges of asset scanning in industrial environments which would typically be even more complex and more connected\textemdash with hundreds of sites holding thousands of OT assets~\cite{DBLP:journals/Dragos18}, both old and new. A typical ICS architecture includes several types of networks:

\begin{figure}
\includegraphics[width=1\columnwidth]{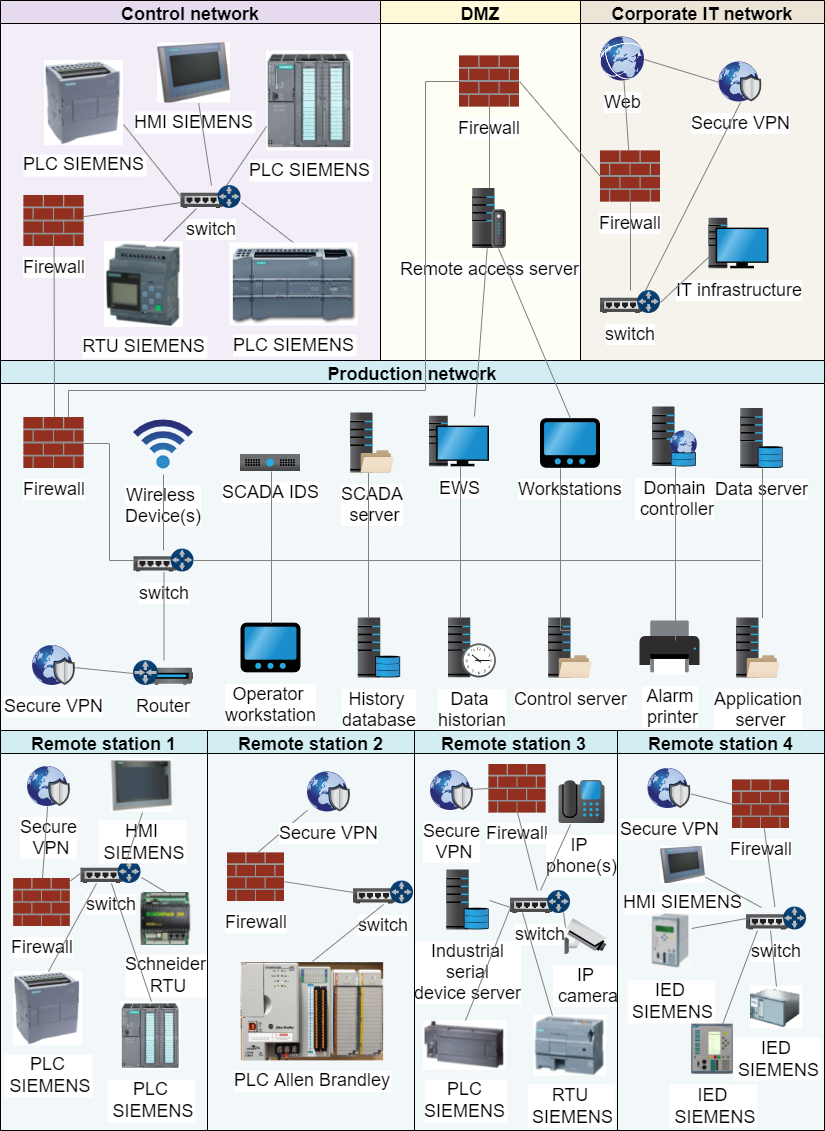}
\caption{ICS/SCADA Architecture}
\label{fig:ICS/SCADA Architecture}
\end{figure}

 \begin{enumerate}
\item Corporate IT network for management of basic plant functions; this includes a demilitarized zone (DMZ) where firewalls filter outgoing or incoming network traffic between corporate and ICS networks.
\item The production network which offers overall monitoring and communication between supervisory control and data acquisition (SCADA) systems and industrial devices \cite{Stouffer2011}. It collects information from field site devices in remote stations and is responsible for controlling the industrial processes through communication links which vary from telephone or power-lines to radio, microwave, cellular and satellite wide area networks. The production network includes SCADA systems, engineering workstations and data historians.
\item The control network provides plant basic functions such as observation and direct control of the manufacturing process. These actions are performed locally by Human Machine Interfaces (HMI), Programmable Logic Controllers (PLC), Remote Terminal Units (RTU) and Intelligent Electronic Devices (IED) from a variety of vendors.
Further, multiple remote stations perform similar functionality and represent smaller remote sites.

\end{enumerate}
Whilst a single zone is likely to utilize the products of a single vendor, different zones within the organization could utilize products from different vendors. ICS devices use a variety of protocols to exchange information. These include both proprietary protocols such as Siemens S7Comm and a range of open protocols such as Modbus, Ethernet/IP and DNP3. In modern times, the majority of these protocols utilize a TCP connection over Ethernet for communication, allowing the systems within the control center to communicate with ICS devices.

\section{Asset scanning challenges for OT Networks}\label{sec:III.background}

\subsection{The challenges posed by industrial/proprietary protocols}
In an OT environment, traditional IT systems such as workstations, servers, switches and routers typically make up only 20\% of the total assets of an industrial plant~\cite{CyberIntegrity17}. Information on these systems can be captured relatively easily using traditional IT asset discovery methods and tools. The other 80\% of OT assets are not as easy to scan for identification or collection of their detailed configuration information because they do not use standard protocols (as noted above). Consequently, a lot of asset owners have limited visibility over their OT assets. Information is often collected manually without the use of scanning tools~\cite{CyberIntegrity17}. This means that the asset information remains static and needs to be manually maintained (leading to the potential for assets being missed). While control system vendors offer tools for scanning their devices, this only partially addresses the problem as industrial environments are typically heterogeneous, involving devices, software and platforms from a variety of vendors as well as legacy and state-of-the-art devices. In Figure \ref{fig:sub-second}, we demonstrate a high-level process of scanning ICS devices and what protocols are applicable in every step. A verbose version of this process can be found in scenario (A) of Table \ref{tab:Scenario (a)} in the appendix, which includes various tool combinations, the challenges asset scanning poses to the devices, and what issues may arise, e.g., device or process disruption, safety compromise, and poor tool performance. 

\vspace{4pt}

\begin{mdframed}[style=MyFrame,nobreak=true,align=center,userdefinedwidth=\columnwidth]
We couldn't find any indication in developers' documentation about the tools we examine in Table~\ref{table:Scanning Tools Overview}, explaining potential issues or impact of using them against live devices. This is the reason why an evaluation of these tools is extremely valuable for practitioners and asset owners, as it offers a map of safe usage examples about asset scanning tools.
\end{mdframed}

\subsection{The challenges of active scanning}
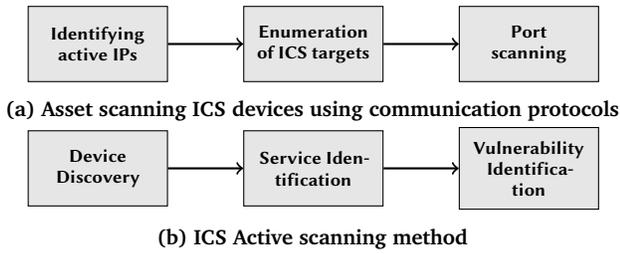
\begin{figure}
\begin{subfigure}{.5\textwidth}
\centering
\begin{tikzpicture}
[node distance = 1cm, auto,font=\footnotesize,
every node/.style={node distance=3cm},
comment/.style={rectangle, inner sep= 5pt, text width=3cm, node distance=0.25cm, font=\scriptsize\sffamily},
force/.style={rectangle, draw, fill=black!10, inner sep=5pt, text width=1.5cm, text badly centered, minimum height=1cm, font=\bfseries\footnotesize\sffamily}] 
\node [force] (rivalry) {Enumeration of ICS targets};
\node [force, left=1cm of rivalry] (suppliers) {Identifying active IPs};
\node [force, right=1cm of rivalry] (users) {Port scanning};
\path[->,thick] 
(suppliers) edge (rivalry)
(rivalry) edge (users);
\end{tikzpicture} 
  \caption{Asset scanning ICS devices using communication protocols}
  \label{fig:sub-second}
\end{subfigure}
\begin{subfigure}{.5\textwidth}
  \centering
\begin{tikzpicture}
[node distance = 1cm, auto,font=\footnotesize,
every node/.style={node distance=3cm},
comment/.style={rectangle, inner sep= 5pt, text width=3cm, node distance=0.25cm, font=\scriptsize\sffamily},
force/.style={rectangle, draw, fill=black!10, inner sep=5pt, text width=1.5cm, text badly centered, minimum height=1cm, font=\bfseries\footnotesize\sffamily}] 
\node [force] (rivalry) {Service Identification};
\node [force, left=1cm of rivalry] (suppliers) {Device Discovery};
\node [force, right=1cm of rivalry] (users) {Vulnerability Identification};
\path[->,thick] 
(suppliers) edge (rivalry)
(rivalry) edge (users);
\end{tikzpicture}
  \caption{ICS Active scanning method}
  \label{fig:sub-third}
\end{subfigure}
\caption{High-level representation of Tables \ref{tab:Scenario (a)} and \ref{tab:Scenario (b)} in the Appendix}
\label{tab:Asset_scanning}
\end{figure}

Asset scanning techniques can either be \textit{passive}~\cite{DBLP:journals/corr/abs-1904-04271}, in which network traffic is collected and the contents of packets analyzed to identify and gain information about hosts, or \textit{active}~\cite{DBLP:journals/scn/CoffeySMJ18}, in which the tool sends a request to a device and analyses the responses. Passive scanning is beneficial as it does not introduce any extra traffic into the network, although this sacrifices accuracy (as only hosts that are actively communicating can be identified). Active scanning can provide a much greater level of accuracy, though can potentially interfere with a target's normal behavior. In Figure \ref{fig:sub-third}, we depict the whole active scanning process and which methods are applicable in every step. More information is available regarding which active tools can reach a specific depth of scanning inside OT networks and can be found in scenario (B), Table \ref{tab:Scenario (b)} in the appendix. Inside a typical IT network, active asset scanning can be achieved through various techniques like ping sweeps or ARP scanning. There are some similarities between ICS and IT devices, such as the use of addressable and routable protocols such as TCP/IP or in particular layers 1, 2 and 3 of the open systems interconnection (OSI) model. Hence, all known ICS scanning tools use the same techniques to scan OT networks. With the exception of some devices that use RS-232 standards for serial communication, the IP addresses of ICS systems can be displayed and analyzed to detect open TCP/UDP ports in the IPv4 address space. Based on the general structure of the ICS network, industrial devices can be accessed via TCP, UDP, and ICMP or ARP scan like IT devices. The usage of active scanning techniques in OT networks poses a risk of industrial process disruption or can put the control devices into stop mode either through a fault caused by a packet going to the wrong port or through resource exhaustion in handling the large volume of traffic such scans often require.
\vspace{2pt}
\begin{mdframed}[style=MyFrame,nobreak=true,align=center,userdefinedwidth=\columnwidth]
There is currently no detailed information for each available asset scanning tool we studied so far on whether it is safe to poll information from OT assets.
\end{mdframed}
\vspace{-7pt}
\subsection{Lack of a common framework to contrast asset scanning tools}
Existing work, largely in industry, has highlighted the importance of continuously monitoring industrial assets in real-time in order to maintain an up-to-date view of the ICS environment~\cite{DBLP:conf/Kongezos13} and promptly detect and report problems or deviations~\cite{DBLP:journals/Cyberx20}. The history of cyberattacks on ICS shows that any kind of disruption of the industrial process can cause severe physical impacts such as environmental damages or even risk loss of life \cite{DBLP:journals/pieee/McLaughlinKWDSM16}. NIST notes that asset management is the first step to run continuously in order to dynamically mitigate cyber security risk \cite{DBLP:journals/Barrett20}. Several case studies have highlighted the need for industrial asset management. Kongezos et al. discuss industrial asset management process and strategies and how they are being implemented in reality using ABB's Asset Optimization software in the Ormen Lange natural gas plant and the Goliat offshore oil field \cite{DBLP:conf/Kongezos13}. Gelle et al. subsequently describe ABB's automation system 800xA and its ability to give a status overview of the control network and a full topology of all connected ICS and IT devices \cite{DBLP:conf/ecbs/GelleKS05}. However, in both the aforementioned cases, it is not clear on which characteristics they rely to choose their asset scanning tools or whether they used a framework to compare their tool features with other scanners.
\vspace{4pt}

\begin{mdframed}[style=MyFrame,nobreak=true,align=center,userdefinedwidth=\columnwidth]
The lack of a common framework to compare asset scanning tools poses a problem for asset owners because they do not know how tool features contrast, hence they do not know which one or which combination is the right one to offer full ICS protocol coverage for heterogeneous ICS systems, fast scanning results for large OT networks or minimum impact on safety and operational reliability.
\end{mdframed}
This can lead to a poor asset inventory for an organization, hence impacting security measures. For example, if deployment-specific information (See Table ~\ref{tab:Asset Scanning depth levels}), is missing from an inventory and an incident occurs, the organization will not be able to quickly restore devices to previous and safe configurations.

\begin{figure*}[!ht]
\includegraphics[width=0.8\textwidth]{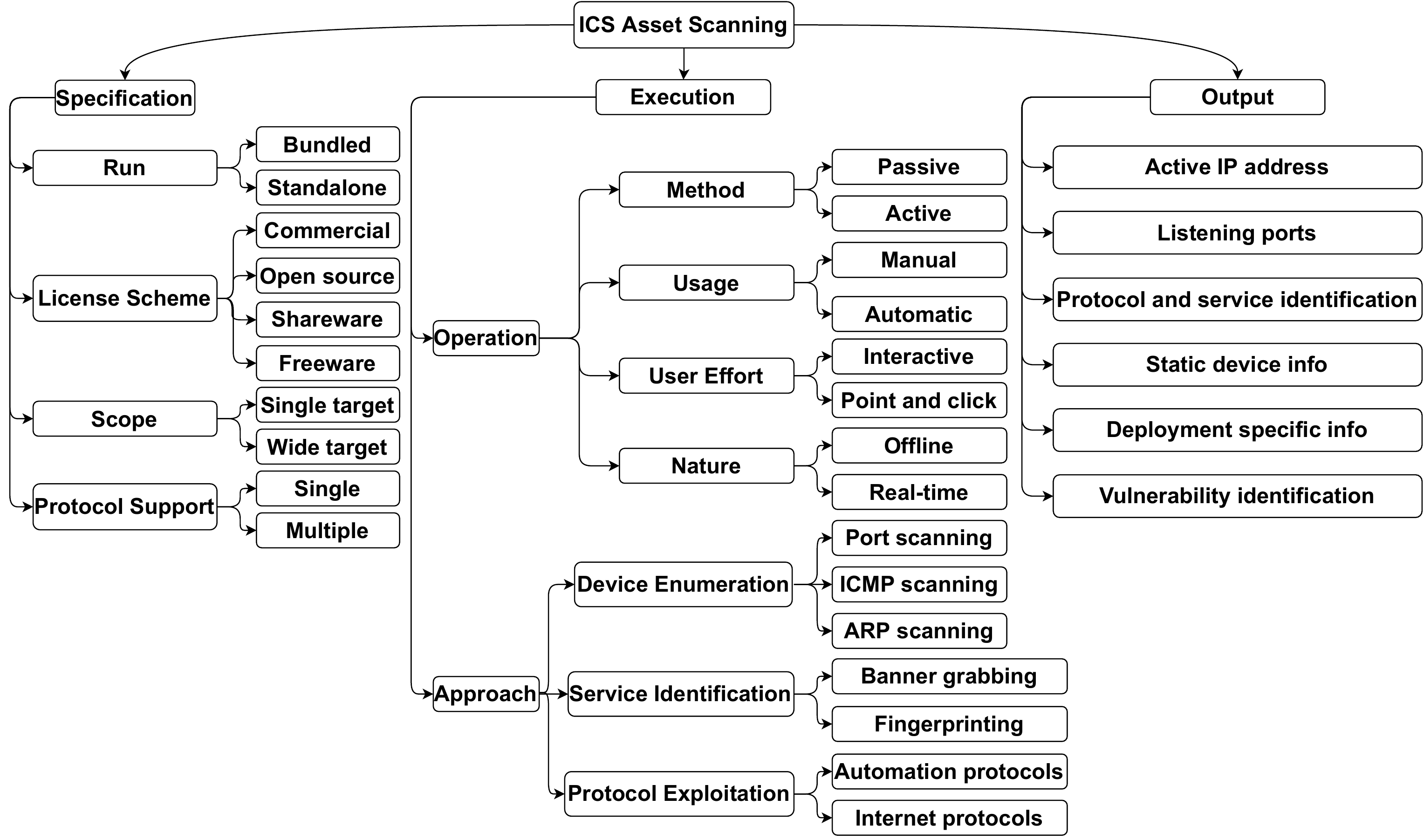}
\centering
\caption{Taxonomy of ICS Asset Scanning Tools}
\label{fig:Taxonomy}
\end{figure*}
\section{Taxonomy of ICS asset scanning features}\label{sec:IV.taxonomy}

Through analysis of various ICS asset discovery tools, we have derived a taxonomy of such tools, which we present in Figure~\ref{fig:Taxonomy}. This taxonomy entails describing, naming, and classifying the various features and steps involved in the scanning process. It is aimed at supporting the planning phase of asset discovery, whereby asset owners can map the scanning features or functionality required to the taxonomy in order to identify the most suitable scanning tools or combination thereof. The taxonomy is derived based on both descriptive information on the wide range of tools available (as provided by developers) and our own insights from practically running the tools and observing their functionality on our testbed (See Section ~\ref{sec:lab}). The taxonomy is divided into three main classes: \textit{Specification}, \textit{Execution} and \textit{Output} with sub-classes capturing more detailed features and functionalities. For each category, we provide an example of a tool in that category. A full list of the tools we studied is available in Table~\ref{table:Scanning Tools Overview}.

\begin{table}[h]
\renewcommand{\arraystretch}{1.25}
\caption{Scanning Tools Overview}
\label{table:Scanning Tools Overview}
\centering
\resizebox{8.3cm}{!}{
\setlength\doublerulesep{0.8pt} 
\begin{tabular}{lccccccccccccccc}
\toprule
\multicolumn{1}{c}{{}} & {\RotText{EtherNet/IP}} & {\RotText{Profinet}}& {\RotText{Profibus}}& {\RotText{Modbus}}& {\RotText{Bacnet}}& {\RotText{S7comm}}& {\RotText{FINS}}& {\RotText{DNP3}}& {\RotText{FF}}& {\RotText{OPC UA}}& {\RotText{SNMP}}& {\RotText{Ethercat}}& {\RotText{HART}}& {\RotText{Version}}& {\RotText{Last Update}} \\ \hline
SIMATIC         &   \xmark         & \OK        & \OK        & \xmark      & \xmark      & \xmark      & \xmark    & \xmark    & \xmark  & \xmark      & \xmark    & \xmark        & \xmark    & v04.00.03   & 08-27-2021     \\ 
\hline
Modscan             & \xmark           & \xmark        & \xmark        & \OK      & \xmark      & \xmark      & \xmark    & \xmark    & \xmark  & \xmark      & \xmark    & \xmark        & \xmark    & v0.1        & 06-01-2021     \\ 
\hline
Nmap                & \OK           & \xmark        & \xmark        & \OK      & \xmark      & \OK      & \OK    & \xmark    & \xmark  & \xmark      & \OK    & \xmark        & \xmark    & v7.9        & 03-10-2020     \\ 
\hline
Plcscan             & \xmark           & \xmark        & \xmark        & \OK      & \xmark      & \OK      & \xmark    & \xmark    & \xmark  & \xmark      & \xmark    & \xmark        & \xmark    & v0.1        & 06-01-2021     \\ 
\hline
Grassmarlin         & \OK           & \xmark        & \xmark        & \OK      & \xmark      & \OK      & \xmark    & \OK    & \OK  & \xmark      & \xmark    & \xmark        & \xmark    & v3.2.1        & 27-06-2017     \\ 
\hline
NetworkMiner        & \xmark           & \xmark        & \xmark        & \OK      & \xmark      & \OK      & \xmark    & \xmark    & \xmark  & \xmark      & \xmark    & \xmark        & \xmark    & v2.7        & 15-06-2021    \\ 
\hline
Sophia              & \OK           & \xmark        & \xmark        & \OK      & \OK      & \OK      & \xmark    & \OK    & \xmark  & \OK      & \xmark    & \xmark        & \xmark    & v3.5        & 03-01-2019     \\ 
\hline
Lansweeper          & \xmark           & \xmark        & \xmark        & \xmark      & \xmark      & \OK      & \xmark    & \xmark    & \xmark  & \xmark      & \OK    & \xmark        & \xmark    & v9.0.10.2 & 30-09-2021     \\ 
\hline
SCADA-CIP & \OK           & \xmark        & \xmark        & \xmark      & \xmark      & \xmark      & \xmark    & \xmark    & \xmark  & \xmark      & \xmark    & \xmark        & \xmark    & v.1.0       & 06-03-2016     \\ 
\hline
Wireshark           & \OK           & \OK        & \xmark        & \OK      & \OK      & \OK      & \xmark    & \OK    & \xmark  & \OK      & \OK    & \OK        & \OK    & v3.4.9      & 06-10-2021     \\ 
\hline
Nessus              & \OK           & \OK        & \xmark        & \OK      & \OK      & \OK      & \xmark    & \OK    & \xmark  & \OK      & \OK    & \xmark        & \xmark    & v8.15.0      & 15-06-2021     \\ 
\hline
OpenVAS            & \OK           & \xmark        & \xmark        & \OK      & \xmark      & \OK      & \OK    & \OK    & \xmark  & \xmark      & \OK    & \xmark        & \xmark    & v21.04.0        & 16-04-2021     \\ 
\hline
scada-tools         & \xmark           & \OK        & \xmark        & \xmark      & \xmark      & \OK      & \xmark    & \xmark    & \xmark  & \xmark      & \xmark    & \xmark        & \xmark    & v1.0        & 26-04-2014     \\ 
\hline
s7scan              & \xmark           & \xmark        & \xmark        & \xmark      & \xmark      & \OK      & \xmark    & \xmark    & \xmark  & \xmark      & \xmark    & \xmark        & \xmark    & v1.03       & 28-12-2018     \\ 
\hline
Redpoint            & \OK           & \xmark        & \xmark        & \xmark      & \OK      & \OK      & \OK    & \xmark    & \xmark  & \xmark      & \xmark    & \xmark        & \xmark    & v1.0          & 08-03-2016     \\ 
\hline
ETTERCAP            & \OK           & \OK        & \xmark        & \OK      & \xmark      & \xmark      & \xmark    & \xmark    & \xmark  & \xmark      & \xmark    & \xmark        & \xmark    & v0.8.3.1      & 01-08-2020    \\ 
\hline
OWASPNettacker  & \OK           & \xmark        & \xmark        & \OK      & \xmark      & \xmark      & \xmark    & \OK    & \xmark  & \xmark      & \xmark    & \xmark        & \xmark    & v2.0        & 12-08-2021     \\ 
\hline
Unicornscan         & \OK           & \xmark        & \xmark        & \xmark      & \xmark      & \xmark      & \xmark    & \xmark    & \xmark  & \xmark      & \xmark    & \xmark        & \xmark    & v0.4.7      & 30-05-2013     \\ 
\hline
nmap-scada          & \xmark           & \xmark        & \xmark        & \xmark      & \xmark      & \OK      & \xmark    & \xmark    & \xmark  & \xmark      & \xmark    & \xmark        & \xmark    & v1.0          & 16-12-2013     \\ 
\hline
icsmaster           & \OK           & \xmark        & \xmark        & \OK      & \OK      & \OK      & \OK    & \OK    & \xmark  & \xmark      & \xmark    & \xmark        & \xmark    & v1.0          & 04-01-2019     \\ 
\hline
Modbusdiscover     & \xmark           & \xmark        & \xmark        & \OK      & \xmark      & \xmark      & \xmark    & \xmark    & \xmark  & \xmark      & \xmark    & \xmark        & \xmark    & v0.3        & 06-09-2018     \\ 
\hline
scadascan           & \xmark           & \xmark        & \xmark        & \OK      & \xmark      & \xmark      & \xmark    & \OK    & \xmark  & \xmark      & \xmark    & \xmark        & \xmark    & v1.0          & 26-10-2011     \\ 
\hline
s7-info             & \xmark           & \xmark        & \xmark        & \xmark      & \xmark      & \OK      & \xmark    & \xmark    & \xmark  & \xmark      & \xmark    & \xmark        & \xmark    & v1.0          & 10-09-2020     \\ 
\hline
plc-scanner         & \xmark           & \OK        & \xmark        & \xmark      & \xmark      & \OK      & \xmark    & \xmark    & \xmark  & \xmark      & \xmark    & \xmark        & \xmark    & v1.2.1        & 06-07-2020     \\ 
\hline
ICS-Hunter          & \OK           & \xmark        & \xmark        & \OK      & \xmark      & \xmark      & \xmark    & \xmark    & \xmark  & \xmark      & \xmark    & \xmark        & \xmark    & v1.0        & 23-03-2020     \\ 
\hline
ModbusScanner       & \xmark           & \xmark        & \xmark        & \OK      & \xmark      & \xmark      & \xmark    & \xmark    & \xmark  & \xmark      & \xmark    & \xmark        & \xmark    & v1.0           & 06-08-2017     \\ 
\hline
ICSY                & \OK           & \xmark        & \xmark        & \xmark      & \xmark      & \xmark      & \xmark    & \xmark    & \xmark  & \xmark      & \xmark    & \xmark        & \xmark    & v1.0          & 19-04-2017     \\ 
\hline
cyberlens           & \OK           & \OK       & \xmark        & \OK      & \xmark      & \OK      & \xmark    & \OK    & \xmark  & \xmark      & \OK    & \xmark        & \xmark    & v1.4        & 03-01-2019     \\
 \toprule
\end{tabular}}
\footnotesize{{\OK} --> applicable | \xmark \ --> non-applicable}\\
\vspace{-8pt}

\end{table}

\subsection{Specification} A range of basic features of a tool -- whether it is standalone or requires other tools to provide asset scanning coverage, its licensing model, scope as well as support for the range of industrial protocols.

\begin{enumerate}
\item Run:
Whether the software can stand on its own or if it comes bundled with another software.
\begin{enumerate}
\item \textit{Bundled}: A tool is not able to run as-is but needs another tool in order to complete the process of scanning, such as scripts. Therefore, these tools are distributed or used with another tool. For example, the \textit{Redpoint tool} is a collection of NSE scripts executed using Nmap.
\item \textit{Standalone}: This category includes software that works alone and is not a part of or uses any bundled software. \textit{SIMATIC} is a standalone software which a user can install as-is to scan Profinet/Profibus or Ethernet networks to identify ICS devices.
\end{enumerate}
\item License Scheme:
The software license category that a tool uses.
\begin{enumerate}
\item \textit{Commercial}: Copyrighted software developed for sale from a company via license or subscriptions, for example \textit{Nessus Professional}.
\item \textit{Open source}: A non-copyrighted software that might be used without restriction and programs with the source code available to everyone. \textit{Plcscan} is a Python script-based software with the source code available to the user for inspection or modification.
\item \textit{Shareware}: Copyrighted software with a trial period after which the user must pay a license fee to continue using it. \textit{Lansweeper} is commercial software with the ability to use a free trial license to test its capabilities for a particular period of time.
\item \textit{Freeware}: The user can use copyrighted software for free without paying any fees. \textit{Grassmarlin} is a free to use tool that provides network scanning of ICS networks available for Windows, Linux Debian and Red-hat distributions without providing the source code to users.
\end{enumerate}
\item Scope:
A tool can scan or identify single or multiple targets inside a network. Therefore, scope specifies whether a tool can target actively or identify passively one or more devices on each run.
\begin{enumerate}
\item \textit{Single target}: This feature allows focused scanning of a single device inside the network. \textit{S7scan} is an active scanning tool that uses the S7 protocol to connect to one PLC on each run to obtain device information.
\item \textit{Wide target}: Wide target scanning tools can identify or scan multiple targets inside a network sub-net or multiple sub-nets. \textit{Sophia} can take as input the whole subnet IP range to analyze network traffic and provides industrial asset identification of all devices inside an ICS network alongside data flow visualization.
\end{enumerate}
\item Protocol Support:
A tool can identify one or multiple industrial communication protocols. Most protocols are specific to ICS devices as vendors manufacture devices to support a single or small set of specific industrial protocols. The industrial protocols identified through experimental scanning of the testbed based on scanners in Table \ref{table:Scanning Tools Overview}, are EtherNet/IP, Modbus, S7comm, SNMP, Profinet. However, except these basic protocols, scanners support also other protocols such as Profibus, Bacnet, FINS, DNP3, FF, OPCUA, Ethercat and HART. The right tool combination can address coverage and compatibility needs inside diverse environments where multiple ICS devices deployed.
\begin{enumerate}
\item \textit{Single}: The tool focuses on a particular target protocol. \textit{SIMATIC} is a SIEMENS proprietary scanning tool able to scan PROFINET/Ethernet networks and identify ICS components such as SIMATIC S7-1200, S7-1500, ET-200s, HMI, SITOP, SCALANCE switches, RFID and MOBY Ident modules.
\item \textit{Multiple}: The tool supports multiple target protocols. \textit{Sophia} is one of these tools, able to identify various free-to-use industrial protocols such as EtherNet/IP, Modbus, Bacnet, S7comm, DNP3 and OPC UA. \\
\end{enumerate}
\end{enumerate}

\subsection{Execution}
The process by which a tool executes a full asset scanning cycle. We split this into two sub-classes: operation and approach.
\begin{enumerate}
\item Operation: The scanning operation is characterized by the following sub-categories.
\begin{enumerate}
\item Method
\begin{enumerate}
\item \textit{Passive scanning}: The tool captures network traffic and extracts information about devices without sending any packets itself. \textit{Ettercap} supports passive identification of several protocols through sniffing live connections, providing information for each target in profile details.

\item \textit{Active scanning}: The tool sends probe packets, targeting hosts inside a network and monitoring their responses to extract information about the devices. A Python-based active scanner is \textit{Plcscan}, able to scan PLC devices using S7comm or Modbus communication protocols.

\end{enumerate}
\item Usage
\begin{enumerate}
\item \textit{Manual}: Tool requires manual user intervention in order to operate, including selecting targets and initiating scanning actions. Such a tool is the python based \textit{Modscan}, where a user is supposed to manually enter a target IP address or sub-net to initiate the scanning process.
\item \textit{Automatic}: Tool automatically scans subnet(s) and returns results to the user, with minimal user intervention past initial configuration. \textit{Sophia} and \textit{Cyberlens} are tools requiring only to set up the adapter for the tool to return info about devices inside a network.
\end{enumerate}
\item User Effort

\begin{enumerate}
\item \textit{Interactive}: The tool requires user interaction to operate, either through an interface or a series of commands within a terminal. \textit{Nmap} is a classic active scanning tool which requires the user to type a set of commands to initiate scanning.
\item \textit{Point and click}: The interface is mainly “Point and click” with none or limited input from the keyboard. \textit{Grassmarlin} is a tool that passively collects information for devices, by analyzing network traffic and the user can acquire them without the need of any further effort.
\end{enumerate}
\item Nature

\begin{enumerate}
\item \textit{Offline}: A software can collect packet capture files of the network traffic and analyze them offline. This refers to a feature some tools, e.g., \textit{Wireshark}, have to analyze a pcap file instead of live sniffing network traffic.
\item \textit{Real-time}: The ability to either actively scan a network for industrial devices or perform live sniffing of network traffic for asset scanning reasons. Any tool performing live interaction with network packets exist in this subclass, such as \textit{Plc-scanner} from Plcdatatools and \textit{Redpoint} script collection.
\end{enumerate}
\end{enumerate}

\item Approach: How current tools could be used on the devices during the scanning phase.
\begin{enumerate}
\item Device Enumeration \\
The process of identifying hosts and the services running on them.
\begin{enumerate}
\item \textit{Port scanning}: Used for probing a whole network or subnet for open ports on the devices. \textit{Lansweeper} is a well-known port scanner to gather details about active IPs.
\item \textit{ICMP scanning}: A simple scanning technique involving a single ICMP ping packet to determine which IP addresses map to live devices. \textit{Nmap} can send ICMP timestamp requests and await ICMP timestamp replies to determine whether a host is alive.
\item \textit{ARP scanning}: Enables the user to discover all the IPv4 network-connected devices through ARP packets. With this technique, IP addresses are mapped to MAC addresses. The “Ping Host” feature of \textit{OpenVAS} can be configured to perform ARP scanning to discover hosts.
\end{enumerate}
\item Service Identification \\
Tool's ability to identify running services based on the targets open ports using the following techniques.
\begin{enumerate}
\item \textit{Banner grabbing}: Can acquire software information revealing insecure and vulnerable applications by sending specially crafted packets to the targets or sniffing the network traffic. One of the features \textit{Unicornscan} holds is the ability to launch asynchronous stateless TCP banner grabbing to gain information about a target.
\item \textit{Fingerprinting}: Is a technique enabling tools to extract information from devices by analyzing packets from targets responses. \textit{Cyberlens} uses fingerprint techniques to analyze various types of packets but also offers customizable fingerprints for identified ports.
\end{enumerate}
\item Protocol Exploitation \\
Network traffic analysis and devices responses to reveal configuration issues or known vulnerabilities based on their communication protocols.
\begin{enumerate}
\item \textit{Automation protocols}: Ability to identify communication protocols used for industrial process automation. \textit{Plcscan} is a dedicated active scanner for scanning devices over S7comm or Modbus protocols only. Wireshark can identify a variety of automation protocols such as EtherNet/IP, Profinet, Modbus, Bacnet, S7comm, DNP3, OPC UA, SNMP, Ethercat and HART.
\item \textit{Internet protocols}: Tools are capable to support various protocols from the internet protocol suite. For this study, we did not include any scanners supporting only internet protocols but many of them can support both internet and industrial protocols. \textit{Grassmarlin} and \textit{Sophia} even-though focused on ICS asset discovery can identify packets from ARP, ICMP, SNMP, SSH, etc. to locate nodes different from ICS devices that usually co-exist inside an OT network.
\end{enumerate}
\end{enumerate}
\end{enumerate}
\subsection{Output}
As ICS scanning tools have distinctive features and capabilities, they offer hence different kinds of output to a user.
\begin{enumerate}

\item{Active IP addresses}: Basic discovery to gain information about active devices inside the ICS network. All tools are able to identify active targets inside an ICS network actively sending packets to hosts or sniffing traffic and perform packet inspection, e.g., Nmap uses ICMP ping to locate active IP addresses within a network.
\item{Listening ports}: Returns a list of open network ports on the device. Next step in asset identification is to determine what TCP/UDP ports are open on the “active” IP addresses. Each open port number determines what protocol is running on the target host. Modscan can scan a whole subnet and return active hosts running Modbus protocol when port 502 is open.
\item{Protocol and service identification}: The tool can identify industrial protocols and/or running services. Wireshark can identify, from deep packet inspection, a variety of automation protocols such as EtherNet/IP, Modbus, S7comm, DNP3 etc.
\item{Static device info}: Identification of device manufacturer or vendor and firmware details or model number. Plcscan, by establishing a connection to an ICS device, can retrieve static info such as manufacturer, firmware and model number.

\item{Deployment specific info}: Retrieve specific, operator set device properties such as Modbus slave ID and module name. S7scan uses, for this purpose, S7comm protocol to connect to PLCs and extract deployment-specific info though “Read SZL” request formats.
\item{Vulnerability identification}: Tool uses collected information from public CVE databases, to present a list of known vulnerabilities affecting a device based on its firmware version. 
\end{enumerate}

\section{Analysis of asset scanning Tools}\label{sec:VI.results}
In this section we map the asset discovery tools to the taxonomy presented in Section~\ref{sec:IV.taxonomy}. This is achieved through analysis of the documentation provided for the tools, and supported through practical analysis on a testbed architecture with real ICS devices. 
\subsection{Testbed Setup}\label{sec:lab}
To test the effectiveness of the different asset discovery tools, we use a testbed similar to remote stations 1 and 2, as seen in Figure ~\ref{fig:ICS/SCADA Architecture}.

\begin{table*}
\centering
\caption{Asset Scanning depth levels}
\label{tab:Asset Scanning depth levels}
\resizebox{17.2cm}{!}{
\begin{tabular}{cccccc}
\hline
\textbf{Level} &
  \textbf{Depth of scanning} &
  \textbf{Scanning properties} &
  \textbf{Exploit potential} &
  \textbf{Consequences} &
  \textbf{Attack examples} \\ \hline
1 &
  IP discovery &
  \begin{tabular}[c]{@{}c@{}}Device IP address\\ is recognizable\end{tabular} &
  \begin{tabular}[c]{@{}c@{}}Identify potential\\  targets\end{tabular} &
  \begin{tabular}[c]{@{}c@{}}Information\\  theft\end{tabular} &
  \begin{tabular}[c]{@{}c@{}}Block information  exchange with\\sensors TCP/UDP, Flooding, Smurf Attack\end{tabular} \\ \hline
2 &
  \begin{tabular}[c]{@{}c@{}}Open ports \\ identification\end{tabular} &
  \begin{tabular}[c]{@{}c@{}}Port scanning\\ return “open”\end{tabular} &
  \begin{tabular}[c]{@{}c@{}}Identify services\\ running in the OS\end{tabular} &
  \begin{tabular}[c]{@{}c@{}}Attacks on\\  workstations\end{tabular} &
  \begin{tabular}[c]{@{}c@{}}Fingerprinting type or version of\\ an open service  HTTP-fingerprinting\end{tabular} \\ \hline
3 &
  \begin{tabular}[c]{@{}c@{}}Protocol \& service\\  identification\end{tabular} &
  \begin{tabular}[c]{@{}c@{}}Identification\\  of industrial\\ protocols\end{tabular} &
  \begin{tabular}[c]{@{}c@{}}Weaponize to\\ exploit specific\\ industrial protocols\end{tabular} &
  \begin{tabular}[c]{@{}c@{}}Information\\ retrieval on ICS\end{tabular} &
  \begin{tabular}[c]{@{}c@{}}Record and replay attack, Tunnel arbitrary\\ traffic over the protocol to evade application\\ layer firewall, Retrieve passwords from \\ traffic using dictionaries, Modify specific \\ packet fields, Identify services running on ports\end{tabular} \\ \hline
4 &
  \begin{tabular}[c]{@{}c@{}}Static device\\  info\end{tabular} &
  \begin{tabular}[c]{@{}c@{}}Retrieve properties:\\  manufacturer,\\  firmware, model number\end{tabular} &
  \begin{tabular}[c]{@{}c@{}}Manually\\ identify CVEs\end{tabular} &
  \begin{tabular}[c]{@{}c@{}}Compromise\\ ICS equipment\end{tabular} &
  \begin{tabular}[c]{@{}c@{}}Device crash, Upload of PLC  memory \\ payload, Remote code  execution\end{tabular} \\ \hline
5 &
  \begin{tabular}[c]{@{}c@{}}Deployment\\ specific info\end{tabular} &
  \begin{tabular}[c]{@{}c@{}}Retrieve properties:\\  Modbus slave ID, module name\end{tabular} &
  \begin{tabular}[c]{@{}c@{}}Modify coils,\\ register values\end{tabular} &
  \begin{tabular}[c]{@{}c@{}}Sabotage or\\ manipulate process\end{tabular} &
  Unauthenticated command execution \\ \hline
6 &
  \begin{tabular}[c]{@{}c@{}}Vulnerability\\  identification\end{tabular} &
  \begin{tabular}[c]{@{}c@{}}Based on properties\\  identification of CVE’s\end{tabular} &
  \begin{tabular}[c]{@{}c@{}}Automatically \\ identify CVEs\end{tabular} &
  \begin{tabular}[c]{@{}c@{}}Stopping of\\  production\end{tabular} &
  \begin{tabular}[c]{@{}c@{}}PLC start/stop/reset, Remote code execution\\ Denial of service, Buffer overflow\end{tabular} \\ \hline
\end{tabular}
}
\footnotesize{Level 1 --> less info | Level 6 --> more info}\\
\end{table*}

\paragraph{\textbf{Remote station 1}}
Remote station 1 contains two PLCs (a Siemens S7-1200 1215c and ET200S), a Siemens KTP1200 Basic HMI, a Schneider SCADAPack32 RTU and a Westermo Industrial Ethernet Switch. The testbed contains devices utilizing the Siemens S7 protocol (the two PLCs and HMI), and Modbus (the SCADAPack32). 
\paragraph{\textbf{Remote station 2}}
In order to also incorporate the commonly used Ethernet/IP protocol, we utilized an architecture similar to remote station 2 in Figure ~\ref{fig:ICS/SCADA Architecture}, which includes an Allen-Bradley ControlLogix 5561 PLC. 

All devices are assigned IP addresses within the same /24 subnet. The scanning machine is also assigned an IP within this same subnet, and connected to a Westermo Switch. In order to test passive analysis tools, the Westermo is configured to mirror all traffic through a dedicated port, which is connected to the laptop through a USB-C to Ethernet adapter. 

\subsection{Asset scanning detail depth}\label{sec:depths}
In order to construct the taxonomy, we extracted all information provided by the tools' developers on the product pages, as well as documentation. Also, we verified and supplemented this through practical testing in our testbed. Asset scanning tools can provide quite different levels of output to the user. Whilst some tools can provide almost all information about a device, other, simpler tools, may only provide one specific piece of information.

Table \ref{tab:Asset Scanning depth levels} is based on our insights from practical experimentation with the tools. Therefore, we define asset scanning depth levels based on the characteristics we have obtained from vendor for each tool and the experimental results from the asset scanning process. Thus, we defined the different scanning depth levels in an order from one, which is the lowest and represent basic information (IP only), to the highest six, which provides verbose information about ICS devices -- a necessary step that leads to vulnerability identification. For every level, we demonstrate the depth each tool can reach, the properties they could retrieve from scanning, the potential exploit a tool can offer to an attacker, the consequences for an ICS network, and some feasible attack examples -- an attacker can launch by utilizing the information extracted by a tool for each level. 

\subsection{Contrasting tools}
Table \ref{table:Scanning Tools Overview} presents information about each tool extracted from documentation and product pages. We summarize information about what automation protocols each tool supports from vendors advertised features, their latest version and last release date. This table also demonstrates the protocol coverage capabilities tools have when deployed inside an OT network. The combined use of documentation and practical testing results allows us to provide a complete mapping as possible. Table \ref{table:Taxonomy mapping to scanning results} depicts the asset scanning results for each tool we evaluated. These results are mapped based on the taxonomy from Figure \ref{fig:Taxonomy} and Table \ref{tab:Asset Scanning depth levels} properties, aiming to show, after practical examination, which technical properties and characteristics these tools actually include. 

For example, many of the smaller open-source tools provide little to no documentation and are simply a GitHub repository with a single file, so practical testing and inspection are required to identify how the tool operates and what communication protocols are supported. Mostly the tools are consistent with what developers and vendors advertise. Almost in all cases except OWASPNettacker, the tools delivered what they promised to do. Although OWASPNettacker claims to identify SCADA devices, it could only discover if the hosts were up and the target's IPs. It is also interesting to mention that 68\% of the tools needed manual input from the user. This mapping is useful to asset owners/operators in identifying which tools may be of use in their particular environments.

The main purpose of open-source and free-to-use asset discovery tools is to provide a way for specialists and security auditors to discover ICS devices and enumerate them to identify if security configuration is missing such as critical firmware updates. However, not all free-to-use asset discovery tools have an active development cycle as we can see from the last update column in Table \ref{table:Scanning Tools Overview}. Also, even though these tools are popular for scanning ICS networks, we cannot define a software’s maturity simply by how long they have been on the market. As a result infrastructure operators may be choosing tools that are not fit for their purpose\textemdash risking disruption to critical infrastructures, or they may choose not to deploy any asset scanning tools but to rely on procurement information they hold about their ICS devices. This leads to gaps in understanding of their assets and vulnerabilities. For the reasons mentioned above, and because there is no guarantee from developers for the safety of using their tools, we chose to test them against real ICS devices.

\vspace{-3pt}
\subsection{Scanning depth levels}
In Table \ref{table:Taxonomy mapping to scanning results} we also illustrate the highest level each scanning tool was able to reach after scanning all industrial targets. Every organization has different needs, so it is not often necessary to launch exhaustive scans against these fragile industrial devices to meet all levels. Utilizing Table \ref{table:Taxonomy mapping to scanning results}, a practitioner can easily identify which tool or combination thereof is applicable and suitable to deploy for their particular requirements. For example, in a use case where we compare active scanning tools supporting purely Modbus protocol, Modscan, Nmap, and ICSY only reach scanning depth level 2 and so only identify which device uses port 502 without providing any further details.

For tools that identify specific vulnerabilities, this is often done by looking up device information in public CVE (Common Vulnerabilities and Exposures) databases. Thomas et al.~\cite{DBLP:journals/Catch20}, showed that information from public CVE databases, in particular, the common platform enumeration (CPE) lists which match devices' firmware to CVEs is not always reliable or accurate. This means that whilst tools will return a list of vulnerabilities, they may miss vulnerabilities if this information is not included. Our proposed taxonomy provides a systematic map of the tools' scanning output capabilities and the kind of information that can be extracted from  ICS devices. As we can observe from table \ref{table:Taxonomy mapping to scanning results}, operators can use such a comparison to focus on only those tools that provide accurate output results such as  static device info and deployment specific info and follow this up by manual validation of these results via CVE databases.

\begin{table*}[h]
\renewcommand{\arraystretch}{0.8}
\caption{Taxonomy mapping to scanning results}
\label{table:Taxonomy mapping to scanning results}
\centering
\resizebox{17.6cm}{!}{
\setlength\doublerulesep{0.8pt} 
\begin{tabular}{clcccccccccccccccccccccccccccccc} \toprule \\[5pt]
 & & {\RotText{SIMATIC}} & {\RotText{Modscan}}& {\RotText{Nmap}}& {\RotText{Plcscan}}& {\RotText{Grassmarlin}}& {\RotText{NetworkMiner}}& {\RotText{Sophia}}& {\RotText{Lansweeper}}& {\RotText{SCADA-CIP}}& {\RotText{Wireshark}}& {\RotText{Nessus}}& {\RotText{OpenVAS}}& {\RotText{scada-tools}}& {\RotText{s7scan}}& {\RotText{Redpoint}}& {\RotText{ETTERCAP}}& {\RotText{OWASPNettacker}}& {\RotText{Unicornscan}}& {\RotText{nmap-scada}}& {\RotText{icsmaster}}& {\RotText{Modbusdiscover}} & {\RotText{scadascan}}& {\RotText{s7-info}}& {\RotText{plc-scanner}}& {\RotText{ICS-Hunter}}& {\RotText{ModbusScanner}}& {\RotText{ICSY}}& {\RotText{cyberlens}}\\ \hline
\multirow{10}{*}{{\RotText{Specification}}} 
 & Bundled & \xmark  & \xmark  & \xmark & \xmark  & \xmark & \xmark & \xmark  & \xmark & \OK  & \xmark  & \xmark & \xmark  & \OK & \xmark & \OK  & \xmark & \xmark & \xmark & \OK  & \OK  & \xmark & \xmark & \OK  & \xmark & \xmark & \xmark & \xmark  & \xmark \\
 & Standalone & \OK  & \OK  & \OK & \OK  & \OK & \OK & \OK  & \OK & \xmark  & \OK  & \OK & \OK  & \xmark & \OK & \xmark  & \OK & \OK & \OK & \xmark  & \xmark & \OK & \OK & \xmark  & \OK & \OK & \OK & \OK  & \OK  \\
 & Commercial & \xmark  & \xmark  & \xmark & \xmark  & \xmark & \xmark & \xmark  & \xmark & \xmark  & \xmark  & \OK & \xmark  & \xmark & \xmark & \xmark  & \xmark & \xmark & \xmark & \xmark  & \xmark & \xmark & \xmark & \xmark  & \xmark & \xmark & \xmark & \xmark  & \xmark  \\
 & Open source & \xmark  & \OK  & \OK & \OK  & \xmark & \OK & \xmark  & \xmark & \xmark  & \xmark  & \xmark & \xmark  & \xmark & \xmark & \xmark  & \xmark & \xmark & \xmark & \xmark  & \xmark & \xmark & \xmark & \xmark  & \xmark & \xmark & \xmark & \xmark  & \xmark  \\
 & Shareware & \OK  & \xmark  & \xmark & \xmark  & \xmark & \xmark & \xmark  & \OK & \xmark  & \xmark  & \OK & \xmark  & \xmark & \xmark & \xmark  & \xmark & \xmark & \xmark & \xmark  & \xmark & \xmark & \xmark & \xmark  & \xmark & \xmark & \xmark & \xmark  & \xmark  \\
 & Freeware & \xmark  & \xmark  & \xmark & \xmark  & \OK & \xmark & \OK  & \xmark & \OK  & \OK  & \xmark & \OK  & \xmark & \xmark & \xmark  & \xmark & \xmark & \OK & \xmark  & \xmark & \xmark & \xmark & \xmark  & \OK & \xmark & \xmark & \xmark  & \OK  \\
 & Single target & \xmark  & \OK  & \OK & \OK  & \xmark & \xmark & \xmark  & \xmark & \xmark  & \xmark  & \OK & \OK  & \xmark & \OK & \OK  & \xmark & \xmark & \OK & \OK  & \OK & \OK & \OK & \OK  & \OK & \OK & \xmark & \OK  & \xmark  \\
 & Wide target & \OK  & \OK  & \OK & \xmark  & \OK & \OK & \OK  & \OK & \OK  & \OK  & \OK & \OK  & \OK & \xmark & \xmark  & \OK & \OK & \OK & \xmark  & \xmark & \xmark & \xmark & \xmark  & \OK & \xmark & \OK & \xmark  & \OK  \\
 & Single & \xmark  & \OK  & \xmark & \xmark  & \xmark & \xmark & \xmark  & \xmark & \OK  & \xmark  & \xmark & \xmark  & \OK & \OK & \xmark  & \xmark & \xmark & \OK & \OK  & \xmark & \OK & \xmark & \OK  & \OK & \OK & \OK & \OK  & \xmark  \\
 & Multiple & \OK  & \xmark  & \OK & \OK  & \OK & \OK & \OK  & \OK & \xmark  & \OK  & \OK & \OK  & \xmark & \xmark & \OK  & \OK & \OK & \xmark & \xmark  & \OK & \xmark & \OK & \xmark  & \xmark & \xmark & \xmark & \xmark  & \OK  \\
 \toprule
\multirow{15}{*}{{\RotText{Execution}}} 
 & Passive & \xmark  & \xmark  & \xmark & \xmark  & \OK & \OK & \OK  & \xmark & \xmark  & \OK  & \xmark & \xmark  & \xmark & \xmark & \xmark  & \OK & \xmark & \xmark & \xmark  & \xmark & \xmark & \xmark & \xmark  & \xmark & \xmark & \xmark & \xmark  & \OK  \\
 & Active & \OK  & \OK  & \OK & \OK  & \xmark & \xmark & \xmark  & \OK & \OK  & \xmark  & \OK & \OK  & \OK & \OK & \OK  & \xmark & \OK & \OK & \OK  & \OK & \OK & \OK & \OK  & \OK & \OK & \OK & \OK  & \xmark  \\
 & Manual & \OK  & \OK  & \OK & \OK  & \xmark & \xmark & \OK  & \OK & \OK  & \xmark  & \xmark & \xmark  & \OK & \OK & \OK  & \xmark & \xmark & \OK & \OK  & \OK & \OK & \OK & \OK  & \OK & \OK & \OK & \OK  & \OK  \\
 & Automatic & \xmark  & \xmark  & \xmark & \xmark  & \OK & \OK & \xmark  & \xmark & \xmark  & \OK  & \OK & \OK  & \xmark & \xmark & \xmark  & \OK & \OK & \xmark & \xmark  & \xmark & \xmark & \xmark & \xmark  & \xmark & \xmark & \xmark & \xmark  & \xmark  \\ 
 & Interactive & \xmark  & \OK  & \OK & \OK  & \xmark & \xmark & \xmark  & \xmark & \OK  & \xmark  & \xmark & \xmark  & \OK & \OK & \OK  & \xmark & \OK & \OK & \OK  & \OK & \OK & \OK & \OK  & \xmark & \OK & \OK & \OK  & \xmark  \\
 & Point and click & \OK  & \xmark  & \xmark & \xmark  & \OK & \OK & \OK  & \OK & \xmark  & \OK  & \OK & \OK  & \xmark & \xmark & \xmark  & \OK & \xmark & \xmark & \xmark  & \xmark & \xmark & \xmark & \xmark  & \OK & \xmark & \xmark & \xmark  & \OK  \\
 & Offline & \xmark  & \xmark  & \xmark & \xmark  & \OK & \xmark & \OK  & \xmark & \xmark  & \OK  & \xmark & \xmark  & \xmark & \xmark & \xmark  & \xmark & \xmark & \xmark & \xmark  & \xmark & \xmark & \xmark & \xmark  & \xmark & \xmark & \xmark & \xmark  & \OK  \\
 & Real-time & \OK  & \OK  & \OK & \OK  & \OK & \OK & \OK  & \OK & \OK  & \OK  & \OK & \OK  & \OK & \OK & \OK  & \OK & \OK & \OK & \OK  & \OK & \OK & \OK & \OK  & \OK & \OK & \OK & \OK  & \OK  \\
 & Port scanning & \xmark  & \OK  & \OK & \OK  & \xmark & \OK & \xmark  & \OK & \OK  & \xmark  & \OK & \OK  & \OK & \OK & \OK  & \xmark & \OK & \OK & \OK  & \OK & \OK & \OK & \OK  & \OK & \OK & \OK & \OK  & \xmark  \\
 & ICMP scanning & \xmark  & \xmark  & \OK & \xmark  & \xmark & \OK & \xmark  & \OK & \xmark  & \xmark  & \OK & \OK  & \OK & \OK & \OK  & \xmark & \OK & \OK & \OK  & \OK & \OK & \xmark & \OK  & \xmark & \xmark & \xmark & \xmark  & \xmark  \\
 & ARP scanning & \xmark  & \xmark  & \OK & \xmark  & \xmark & \xmark & \xmark  & \OK & \xmark  & \xmark  & \OK & \OK  & \OK & \xmark & \xmark  & \xmark & \xmark & \OK & \xmark  & \xmark & \xmark & \xmark & \xmark  & \xmark & \xmark & \xmark & \xmark  & \xmark  \\
 & Banner grabbing & \xmark  & \xmark  & \OK & \xmark  & \xmark & \xmark & \xmark  & \xmark & \xmark  & \xmark  & \OK & \OK  & \xmark & \xmark & \xmark  & \xmark & \xmark & \OK & \xmark  & \xmark & \xmark & \xmark & \xmark  & \xmark & \xmark & \xmark & \xmark  & \xmark  \\
 & Fingerprinting & \xmark  & \xmark  & \OK & \xmark  & \OK & \OK & \OK  & \OK & \xmark  & \OK  & \OK & \OK  & \xmark & \xmark & \xmark  & \OK & \xmark & \OK & \xmark  & \xmark & \xmark & \xmark & \xmark  & \xmark & \xmark & \xmark & \xmark  & \OK  \\
 & Automation protocols & \OK  & \OK  & \OK & \OK  & \OK & \OK & \OK  & \OK & \OK  & \OK  & \OK & \OK  & \OK & \OK & \OK  & \OK & \OK & \OK & \OK  & \OK & \OK & \OK & \OK  & \OK & \OK & \OK & \OK  & \OK  \\
 & Internet protocols & \xmark  & \xmark  & \OK & \xmark  & \OK & \OK & \xmark  & \OK & \xmark  & \OK  & \OK & \OK  & \xmark & \xmark & \xmark  & \OK & \xmark & \OK & \xmark  & \xmark & \xmark & \xmark & \xmark  & \xmark & \xmark & \xmark & \xmark  & \xmark  \\
 \toprule
\multirow{6}{*}{{\RotText{Output\slash Scanning Depth}}} 
 & 1 Active IP address & \OK  & \OK  & \OK & \OK  & \OK & \OK & \OK  & \OK & \OK  & \OK  & \OK & \OK  & \OK & \OK & \OK  & \OK & \OK & \OK & \OK  & \OK & \OK & \OK & \OK  & \OK & \OK & \OK & \OK  & \OK  \\
 & 2 Listening ports & \OK  & \OK  & \OK & \OK  & \xmark & \OK & \xmark  & \xmark & \OK  & \OK  & \OK & \OK  & \OK & \OK & \OK  & \OK & \xmark & \OK & \OK  & \OK & \OK & \xmark & \OK  & \xmark & \OK & \xmark & \OK  & \xmark  \\
 & 3 Protocol and service identification & \OK  & \xmark  & \xmark & \OK  & \OK & \xmark & \OK  & \OK & \OK  & \OK  & \OK & \OK  & \OK & \OK & \OK  & \OK & \xmark & \OK & \OK  & \OK & \OK & \OK & \OK  & \xmark & \OK & \OK & \xmark  & \OK  \\
 & 4 Static device info & \OK  & \xmark  & \xmark & \OK  & \OK & \OK & \OK  & \OK & \OK  & \xmark  & \OK & \OK  & \OK & \OK & \OK  & \OK & \xmark & \xmark & \OK  & \xmark & \xmark & \xmark & \OK  & \OK & \OK & \xmark & \xmark  & \OK  \\
 & 5 Deployment specific info & \xmark  & \xmark  & \xmark & \OK  & \xmark & \xmark & \xmark  & \xmark & \xmark  & \xmark  & \xmark & \OK  & \xmark & \OK & \OK  & \xmark & \xmark & \xmark & \xmark  & \OK & \OK & \OK & \OK  & \OK & \xmark & \xmark & \xmark  & \xmark  \\
 & 6 Vulnerability identification & \xmark  & \xmark  & \xmark & \xmark  & \xmark & \xmark & \xmark  & \xmark & \xmark  & \xmark  & \OK & \OK  & \xmark & \xmark & \xmark  & \xmark & \xmark & \xmark & \xmark  & \xmark & \xmark & \xmark & \xmark  & \xmark & \xmark & \xmark & \xmark  & \xmark     \\
\toprule
\end{tabular}}
\footnotesize{{\OK} --> applicable | \xmark \ --> non-applicable}\\
\end{table*}

\subsection{Insights into specific tools}
Based on Table \ref{table:Scanning Tools Overview} and Table \ref{table:Taxonomy mapping to scanning results}, we can deliver useful insights on specific tools, before they are used against "live" systems.

\subsubsection{Practical scanning results} 
Scanning with scadascan didn't return static device info (Level 4) but only the first unit ID (Level 5) as expected and Lansweeper was able to understand that a control system was behind an IP address but without deployment-specific information about it, justifying why it can reach only Level 4. Modbusdiscover and icsmaster returned information of SID, slave ID data and device MAC address as expected — that is why Level 4 is out of reach for them.

There are also passive scanning tools with real-time sniffing and fingerprinting capabilities able to perform off-line pcap analysis to identify ICS devices. Accordingly, cyberlens, Sophia, GRASSMARLIN, NetworkMiner, Wireshark and ETTERCAP were able to identify only the one device in the testbed that uses Modbus protocol (e.g. Table \ref{table:Scanning Tools Overview}). Nmap is quite easy to use, able to return open ports (102/tcp open iso-tsap), MAC addresses and vendor for the S7-1200, ET 200S and HMI. In contrast, Nmap based tools using embedded scripts can return verbose results for each target. s7-info is able to return information about module number, hardware number, version, system name and vendor satisfying Level 1 to Level 5 of scanning depth. Additionally, PlcScan uses active scanning and can provide even more detailed results, including information like PLCs firmware version, plant identification, name and a serial number of the module (Level 1 to Level 5) but only for ET 200S not for S7-1200. Nmap and PlcScan were not able to provide the same amount of information for the newer PLC S7-1200 as they did with the older ET200S and for the HMI, as it does not use the S7 protocol. Likewise, GRASSMARLIN was able to provide verbose information using passive pcap analysis for the ET 200S but not adequate information about the S7-1200 and HMI. It is worth mentioning, that GRASSMARLIN can depict a network graph, something that can help cyber security engineers to acquire deep knowledge of all existing components inside their networks. 

Only basic information such as device type, article number, firmware version and MAC addresses were retrieved by testing SIMATIC tool. Results are as expected, as SIMATIC is used to create a mapping of all accessible devices on the network, to use with the other Siemens tools. The rest of the tools follow in the same pattern which means verbose information about ET 200S which is an older model and less or essential information about the newer model S7-1200 and Siemens HMI. s7scan was able to return enough details even for S7-1200 (Level 5) but nothing for the HMI device, while Unicornscan can only identify the iso-tsap protocol on targets (Level 3) in contrast with OWASPNettacker (Level 1) that only found alive nodes through an icmp scan.

\subsubsection{Issues of active scanning}
During Nmap scanning against one of the PLCs, we discovered a potential vulnerability. Usage of certain flags against a PLC device causes the device to enter into a failure state. As a result, the device's LEDs begun flashing and required a full power circle to restore the device to a working state. We reported this potential vulnerability because the issue could be reproduced accidentally by someone using the Nmap tool in a standard way with no specialist skill requirements. The vendor was unable to replicate the result and responded that the device has end-of-life. However, the device is widely deployed in the industry and is still available for sale through third party sellers. Users can find the taxonomy useful to understand which tools may pose risks due to active scans depending on the age of the infrastructure and the specific devices deployed. 

There are also some tools inside the list used mainly as vulnerability scanners but since they use asset scanning techniques we can enlist them as well. OpenVAS could identify S7-1200 and the RTU but couldn't enumerate the HMI even though it could identify the 102 tcp port which indicates a Siemens device. During this basic testing with OpenVAS, we discovered another failure caused by scanning behavior, which is the subject of ongoing investigation and responsible disclosure. One of the devices enters a failure state (requiring a physical intervention of power cycling the device to recover) when exposed to a particular scanning technique  utilized by multiple tools in our testing. When in this state, the device becomes inoperable.

\subsubsection{Discussion}
The knowledge of the scanning depth and the quality of information each tool can reach is based on the scanning levels we defined in Table \ref{tab:Asset Scanning depth levels}. Asset owners can use the taxonomy as a basis to compare various tools they may be considering, the exact results they should expect to see after scanning, and understand whether a tool poses any risk for the device or the industrial process. The taxonomy mapping in Table \ref{table:Taxonomy mapping to scanning results} provides a clear exemplar of such contrast and the comparative analysis the taxonomy enables. 

Practitioners can use such a comparative analysis to narrow their focus only on tools that perform passive network scanning or pcap analysis -- methods that pose no risk to the ICS device's operation. They may also use the the quality of information (based on the scanning depths) these tools can reach as further criteria to refine the list. They can also identify which tools are ICS-specific and which ones operation across IT and OT networks -- and depending on the use case may focus on a specialized tool for ICS or one that can span their IT/OT infrastructure.  

Our analysis also highlights the need for more specialized tools to support more ICS communication protocols whether proprietary or not, as well as the need to enhance tools' fingerprinting capabilities to support more types of ICS devices and offer a better quality of scanning including more device information in the results. 

As noted above, our experimental analysis is a baseline against which other tools and future versions of these tools can be compared and evaluated accordingly. Finally, we highlight that more specialized vulnerability scanners are needed, with a focus on ICS networks and their critical properties such as safety, reliability, and robustness.

\section{Related work}\label{sec:VI.related work}
Kyle Coffey et al.\cite{DBLP:journals/scn/CoffeySMJ18}, aim at identifying the way asset scanning tools interact with ICS devices and whether they are able to cause any kind of disruption to the process. Rodofile et al.\cite{DBLP:conf/acsc/RodofileRF16} focus solely on the discovery of DNP3 devices inside an industrial network. They developed a technique to identify DNP3 masters and slaves, in addition to ARP and port scanning using Nmap functionality for a given address space. Myers et al. provide a taxonomy of internet scanning tools that identify exposed ICS devices to the Internet. They include a tools comparison (Zmap, Masscan, Unicornscan), including their properties and capabilities such as scan method, packet transmission, etc.~\cite{DBLP:conf/auisc/MyersFR15}. They also discuss two stages for Internet-wide scanning, where the first identifies target IPs and the second queries known ports for running services. These two stages are similar to our Level 1 and 2 depths of scanning, even though we focus on local network scanning only. 

In our work, we create for users an industrial asset scanning guide that also includes information about which tools are safe and appropriate for their environments. We contrast tools' capabilities to identify static and deployment specific properties through experimental analysis for a variety of industrial automation protocols. Also, to our knowledge, we are the first to introduce a taxonomy in Figure \ref{fig:Taxonomy} for ICS scanning tools to contrast their characteristics. We evaluate all the available scanning stages ranging from level 1 to level 6 in Table \ref{tab:Asset Scanning depth levels}, including the expected outcome, the potential attacks, and consequences. Throughout evaluation in our testbed, we demonstrate in Table \ref{table:Taxonomy mapping to scanning results} the tools' real features and capabilities. This experimental analysis is a baseline to which other asset scanning tools can be compared efficiently.

\section{Conclusion and future work}\label{sec:VII.conclusion}
We have presented a practical evaluation in order to demonstrate our mapping of tools to the proposed taxonomy. The taxonomy offers -- to the research community and asset owners -- a common means to contrast scanning tools. Such a comparative analysis imparts an understanding of the tools' potential to provide adequate coverage (in terms of asset identification) and also their potential for disruption to critical processes. This experimentation is a first step in this regard. We are in the process of performing a more detailed evaluation of these tools in a much more complex environment, utilizing the full OT network and a wider range of devices. As well as measuring the effectiveness of the tools under different network topologies, we will closely monitor the devices for any identifiable negative behavior introduced through such scanning. This monitoring will include the use of an industrial physical process to identify effects on the process.

\bibliographystyle{ACM-Reference-Format}
\bibliography{references.bib}
\onecolumn
\appendix
\section*{Appendix}\label{appendix:raw}

\begin{table*}[hbp]
  \vspace*{1 cm}
  \centering
  \caption{Scenario (a)}
  \label{tab:Scenario (a)}
  \resizebox{18cm}{!}
  { 
\begin{tabular}{lll} 
\hline
\multicolumn{3}{c}{\textbf{Asset scanning ICS devices using communication~ protocols }}                                                                                                        \\ 
\hline
\multicolumn{1}{c}{\tikzmark{a}{\textbf{Identifying~active IPs}}}                                                                                                   & \multicolumn{1}{c}{\tikzmark{b}{\textbf {Enumeration~of ICS targets}}}                                                                                         & \multicolumn{1}{c}{\tikzmark{c}{\textbf {Port~scanning}}}                                                                                        \\ 
\hline
\begin{tabular}[c]{@{}l@{}}
\textbf{ICMP:~}Nmap,\\ NetworkMiner, Lansweeper,\\ Nessus, OpenVAS,\\ scada-tools, s7scan,\\ Redpoint, \\OWASPNettacker, \\Unicornscan, nmap-scada, \\icsmaster, Modbusdiscover,\\s7-info\\ \textbf{ARP}: Nmap, Lansweeper, \\Nessus, OpenVAS, \\scada-tools,Unicornscan\end{tabular} & \begin{tabular}[c]{@{}l@{}}

\textbf{EtherNet/IP}: Nmap, SCADA-CIP, \\Nessus, OpenVAS, Redpoint,\\OWASPNettacker, Unicornscan, icsmaster, \\ICS-Hunter, ICSY\\\textbf{Profinet}: scada-tools, Nessus, plc-scanner\\\textbf{Profibus}: N/A\\\textbf{Modbus}: Modscan, Nmap, Plcscan, \\NetworkMiner, Nessus, OpenVAS, \\OWASPNettacker, icsmaster, \\modbus-discover, scadascan, ICS-Hunter,\\ModbusScanner\\\textbf{Bacnet}: Nessus, Redpoint, icsmaster\\\textbf{S7comm}: Nmap, Plcscan, NetworkMiner, \\Nessus, scada-tools, s7scan, Redpoint, \\icsmaster, s7-info, plc-scanner, \\lansweeper, OpenVAS\\\textbf{FINS}: Nmap, OpenVAS, Redpoint, icsmaster\\\textbf{DNP3}: Nessus, OpenVAS, OWASPNettacker,\\icsmaster, scadascan\\\textbf{FF}: N/A\\\textbf{OPC UA}: Nessus~\\\textbf{SNMP}: Lansweeper, Nessus, OpenVAS, Nmap\\\textbf{Ethercat}: N/A~\\\textbf{HART}: N/A\end{tabular} & \begin{tabular}[c]{@{}l@{}}\textbf{EtherNet/IP}: Nmap, SCADA-CIP, \\ Wireshark, Nessus, OpenVAS, Redpoint, \\ ETTERCAP, OWASPNettacker,\\ Unicornscan, nmap-scada,\\ icsmaster, ICS-Hunter,~ICSY\\\textbf{Profinet}: SIMATIC, Nessus,\\ scada-tools, ETTERCAP, plc-scanner\\\textbf{Profibus}: SIMATIC~\\\textbf{Modbus}: Modscan, Nmap, Plcscan, \\NetworkMiner, Wireshark, Nessus, \\OpenVAS, ETTERCAP, OWASPNettacker,\\icsmaster, Modbus-discover, ICS-Hunter, \\ModbusScanner\\\textbf{Bacnet}: Wireshark, Nessus, Redpoint, \\icsmaster\\\textbf{S7comm}: Nmap, Plcscan, NetworkMiner, \\Wireshark, scada-tools, s7scan, Redpoint, \\nmap-scada, icsmaster, s7-info, plc-scanner, \\Lansweeper\\\textbf{FINS}: Nmap, OpenVAS, Redpoint, icsmaster\\\textbf{DNP3}: Wireshark, Nessus, OpenVAS, \\OWASPNettacker,~icsmaster, scadascan\\\textbf{FF}: N/A\\\textbf{OPC UA}: Wireshark, Nessus\\\textbf{SNMP}: Lansweeper, Wireshark, Nessus, \\OpenVAS~\\\textbf{Ethercat}: Wireshark\\\textbf{HART}: Wireshark\end{tabular}  \\ 
\hline
\multicolumn{1}{c}{\textbf{}}                                                               & \multicolumn{1}{c}{\textbf{Issues}}                                                                  & \multicolumn{1}{c}{\textbf{}}                                                                                                  \\ 
\hline
\begin{tabular}[c]{@{}l@{}}Not all tools are safe to use against\\fragile devices~as no assurance or \\instructions exist from vendors.\\ Disrupt the process and cause a\\denial of service~to devices due to\\lack of IP network stack robustness. \end{tabular}                                                                                                    & \begin{tabular}[c]{@{}l@{}}Some communication protocols are not \\covered by scanning tools.\\ Not clear what impact could be on safety from \\the use of some or combination of tools. \end{tabular}                                                                                                          & \begin{tabular}[c]{@{}l@{}}Some communication protocols are not \\covered by scanning tools. \\ Even though a toolset can be defined to \\cover a heterogeneous network, still not \\clear which~tools to use. \\ Disrupt the process and cause a denial \\of service. \end{tabular}                                                                                \\ 
\hline

\hline
\end{tabular}
\link{a}{b}
\link{b}{c}
}
\end{table*}

\begin{table*}[t]
  \centering
  \caption{Scenario (b)}
  \label{tab:Scenario (b)}
  \resizebox{18cm}{!}
  { 
\begin{tabular}{ccc} 
\hline
\multicolumn{3}{c}{\textbf{ICS Active scanning method}}                                                                                                        \\ 
\hline
\tikzmark{a}{\textbf{Device~Discovery}}                                                                                      & \tikzmark{b}{\textbf{Service~Identification}}                                                        & \tikzmark{c}{\textbf{Vulnerability Identification}}                                                                                             \\ 
\hline
\multicolumn{3}{l}{\begin{tabular}[c]{@{}l@{}}\textit{Active method generates network traffic querying devices but returns more information about assets than passive method. }\end{tabular}}                                                                           \\ 
\hline
\multicolumn{1}{l}{\begin{tabular}[c]{@{}l@{}}\textbf{Scanning approach:~}Port scanning, \\ICMP scanning, ARP scanning\\\textbf{Outcome}: Identification of active ports and open\\ports using syn scans, ping sweeps and arp scan\\\textbf{Active tools}: Modscan, Nmap, Plcscan, \\Lansweeper, SCADA-CIP, Nessus,~\\OpenVAS, scada-tools, s7scan, Redpoint,\\OWASPNettacker,\\Unicornscan, nmap-scada, icsmaster, \\Modbusdiscover, scadascan, s7-info,\\plc-scanner, ICS-Hunter, ModbusScanner, ICSY\end{tabular}} & \multicolumn{1}{l}{\begin{tabular}[c]{@{}l@{}}\textbf{Scanning approach:~}Banner grabbing,\\Fingerprinting\\\textbf{Outcome}: Identification of industrial protocols,\\Operating systems~and services,\\Static or deployment specific device info\\\textbf{Active tools}:~Nmap, Lansweeper, Nessus,\\ OpenVAS, Unicornscan, Plcscan,\\ Lansweeper, SCADA-CIP,\\scada-tools,~s7scan, Redpoint, Unicornscan, \\nmap-scada, icsmaster, Modbusdiscover, \\scadascan, s7-info, plc-scanner, ICS-Hunter, \\ModbusScanner, SIMATIC~\end{tabular}} & \multicolumn{1}{l}{\begin{tabular}[c]{@{}l@{}}\textbf{Scanning approach:}\\Automation protocols\\Internet protocols\\\textbf{Outcome}:\\Missing devices patches\\Unnecessary running \\services\\Configuration issues\\\textbf{Active tools}:~\\Nessus, OpenVAS\end{tabular}}  \\ 
\hline
\textbf{}                                                                                                                                                                                                        & \textbf{Issues}                                                                                                                                                                                                                                & \textbf{}                                                                                                                                                                 \\ 
\hline
\multicolumn{1}{l}{\begin{tabular}[c]{@{}l@{}}Scan is not running continuously hence \\cannot detect transient IPs or listen to \\only devices. \end{tabular}}                                                                                                                                                                                                                                                                                                                                                                               & \multicolumn{1}{l}{\begin{tabular}[c]{@{}l@{}}Active scanning is faster than passive but by \\sending packets to query ICS devices increase\\the risk of disruption with either incompatible\\queries or with increased network traffic to a\\network.\end{tabular}}                                                                                                                                                                                                                                                                                              & \multicolumn{1}{l}{\begin{tabular}[c]{@{}l@{}}Incompatible queries may lead\\to disruption.\\Live testing to verify vulnerabilities\\identified in the scan is a challenge.\end{tabular}}                                                                                                                                \\
\hline
\end{tabular}
\link{a}{b}
\link{b}{c}
}
\end{table*}

\end{document}